# Visual DCT - Visual EPICS Database Configuration Tool

M. Sekoranja, Jozef Stefan Institute (JSI), Ljubljana, SI 1000, Slovenia
S. Hunt, Paul Scherrer Institute (PSI), Villigen, CH 5232, Switzerland
A. Luedeke, Paul Scherrer Institute (PSI), Villigen, CH 5232, Switzerland

Abstract

Visual DCT [1] is an EPICS configuration tool completely written in Java and therefore supported in various systems. It was developed to provide features missing in existing configuration tools as Capfast [2] and GDCT [3]. Visually Visual DCT resembles GDCT - records can be created, moved and linked, fields and links can be easily modified. But Visual DCT offers more: using groups, records can be grouped together in a logical block, which allows a hierarchical design. Additionally indication of data flow direction using arrows makes the design easier to understand. Visual DCT has a powerful DB parser, which allows importing existing DB and DBD files. Output file is also DB file, all comments and record order is preserved and visual data saved as comment, which allows DBs to be edited in other tools or manually. Great effort has been taken and many tricks used to optimize the performance in order to compensate for the fact that Java is an interpreted language.

## 1 BASIC PRINCIPLES

VisualDCT is designed to create and maintain EPICS record instance database (.db) files. In order for VisualDCT to execute properly, a database definition (.dbd) file has to be provided which contains the specifications for the various record and device types that they intend to reference in any record instance database (.db) file to be created by VisualDCT. Once a database definition (.dbd) file has been specified, records can be created, copied, renamed, etc. using the various facilities provided by the VisualDCT.

As the user interacts with the various VisualDCT windows, selections, and data entry fields, the results of these interactions are displayed on the screen. Revisions and data entry updates of record instance data displayed on the screen do not replace previously stored record instance data until the user saves currently modified record instance database (.db) file. As VisualDCT executes, it attempts to trap and display the most common situations that might lead to diminishing the integrity of the user supplied information.

In order to run VisualDCT, Java Runtime Environment 2 [4] is needed. VisualDCT is distributed as a Java ARchive package (.jar file), so there is only one file in the distribution.

## 2 FEATURES

### 2.1 Rapid Database Development (RDD)

VisualDCT can be considered as a rapid database development tool - intuitive database construction can be done quickly with a few simple mouse-clicks minimizing all unnecessary keyboard input as in ordinary text editors. Visualization of the record instance database makes databases easier to understand, errors are much easier to find (e.g. broken links are indicated by a red cross) and helps find a better design of the databases, allowing to use hierarchical design and split databases into logical blocks.

### 2.2 Database file parser, input/output file

VisualDCT creates and maintains only one file, the record instance database (.db) file, and does not have any additional graphical information file avoiding any possible consistency problems when having multiple files. All necessary visual composition data is stored as a comment at the end of the record instance database (.db) file.

An example of simple record instance (.db) file:

```
#! Generated by VisualDCT for Java v2.0

# this is a record comment
record(ai,ai001) {
  # this is a field comment
  field(INP,"ao001")
}

# another record comment
record(ao,ao001) {
}

#! Further lines contain layout data used by VisualDCT
#! Record(ai001,2241,2345,0,1,"ai001")
#! Field("ai001.INP",16711731,1,"ai001.INP")
#! Link("ai001.INP","ai001/INP")
#!
Connector("ai001/INP","ao001.VAL",2505,2495,0,"")
#! Record(ao001,2641,2500,0,1,"ao001")
```

```
#! Field("ao001.VAL",16711731,0,"ao001.VAL")
```

VisualDCT has a powerful parser which has the ability to parse already existing record instance database (.db) files, files which have been created or modified with other tools. It also detects syntax errors in databases. Defective visual composition data or their absence are safely handled and do not raise any critical error. VisualDCT just automatically layouts all objects that have no visual data. What is more, VisualDCT preserves comments and record/field order in the record instance database, which offers the ability to edit your databases in other tools or text editors without making any harm to the databases and VisualDCT.

## 2.3 Group, record, link representation

A record is represented as a rectangle with its type and name written inside. Below the line inside the record there is an area where all fields with non-default value are shown.

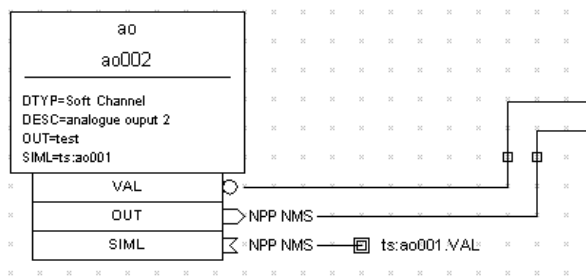

Figure 1 – Representation of the record

There are three types of fields that can appear as part of the record (squares below the record): VARIABLE (data), INPUT, OUTPUT (and FORWARD) fields. Variable fields hold a piece of data, such as the VAL or HIHI fields. Since the variable fields can be populated by other record's output fields and read by other record's input fields, a field node will appear below the record. Additionally indication of data flow direction using arrows makes fields easy to distinguish: circle for VARIABLE fields, out-arrow for OUTPUT and FORWARD fields and in-arrow for INPUT fields. A broken line can be drawn between any two linkable fields simply by adding connectors (moveable small squares on a link line). If a link is an inter-group link (link between two fields which are not in the same group), the link is represented as a line going in the screen with the target link name shown by side.

A group is represented as a rectangle with its name inside.

## 2.4 Grouping - hierarchy support

VisualDCT supports hierarchical design of the databases. Hierarchy is based on the naming, for instance record with name *grp1:ao001* belongs to group *grp1* and record *grp1:grp2:ao002* belongs to group *grp2* which belongs to *grp1*, so groups can be also nested. In previous examples the ":" character was used as a grouping separator, which is the default, but it can be easily set in the preferences menu.

Double click on the group goes (zooms) into the group and shows only the records and groups in this group.

## 3 USER INTERFACE

Like each powerful IDE, also VisualDCT provides indispensable facilities such as clipboard and undo support, too. Great effort was given to synchronization between the record instance database and its visualization. Every change done visually is immediately reflected in the database and vice versa; all actions like moving, renaming and deletion of records which affect links are automatically fixed by the VisualDCT, since all link connections are interpreted in real-time.
(See Figure 2.)

## 4 FUTURE PLANS

Since VisualDCT is still an active project, there are some features to be implemented in the future releases of VisualDCT:
- XML support
- Add arbitrary text, lines, rectangles for documentation
- Handling more than one DB at the same time
- Wrapper to handle so-called startup scripts
- Optimize print preview performance
- Debugging (with active connection to EPICS DB)
- Alternative hierarchy support (like in Capfast)
- CSML [6] support
- … and any user feature requests

We are waiting for user feedback to set priorities.

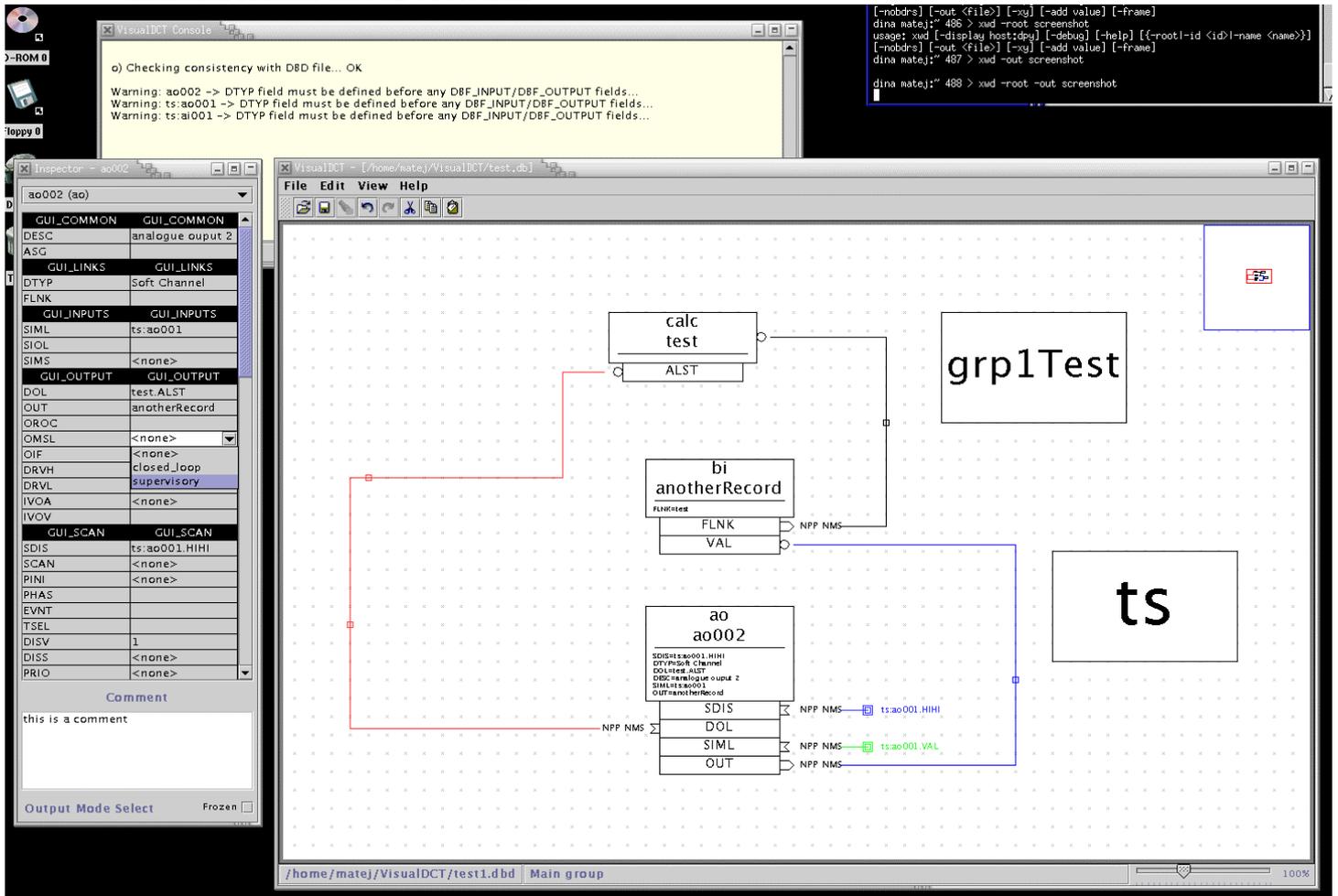

Figure 2 – Graphical User Interface of the VisualDCT tool.